\documentclass{llncs}
\usepackage{graphicx}
\usepackage{caption}
\usepackage{subcaption}
\usepackage{amssymb}
\usepackage{amsmath}

\usepackage[font={small}]{caption, subfig}

\begin{document}

\mainmatter              
\title{From ``I love you babe'' to ``leave me alone'' - Romantic Relationship Breakups on Twitter}
%

\author{Venkata Rama Kiran Garimella\inst{1} \and Ingmar Weber\inst{2} \and
Sonya Dal Cin\inst{3}}
\institute{Aalto University, \email{kiran.garimella@aalto.fi}
\and
Qatar Computing Research Institute, \email{iweber@qf.org.qa}
\and
University of Michigan, \email{sdalcin@umich.edu}
}

\maketitle              

\begin{abstract}
We use public data from Twitter to study the breakups of the romantic relationships of 661 couples. Couples are identified through profile references such as @user1 writing ``@user2 is the best boyfriend ever!!''.
Using this data set we find evidence for a number of existing hypotheses describing psychological processes including
(i) pre-relationship closeness being indicative of post-relationship closeness,
(ii) ``stonewalling'', i.e., ignoring messages by a partner, being  indicative of a pending breakup, and
(iii) post-breakup depression. 
We also observe a previously undocumented phenomenon of ``batch un-friending and being un-friended'' where users who break up experience sudden drops of 15-20 followers and friends.

Our work shows that public Twitter data can be used to gain new insights into psychological processes surrounding relationship dissolutions, something that most people go through at least once in their lifetime.

\keywords{relationships; breakups; Twitter}
\end{abstract}

\section{Introduction}

The breakup of a romantic relationship is one of the most distressing experiences one can go through in  life. It is estimated that more than 85\% of Americans~\cite{battaglia1998breaking} go through this process at least once in their life time. Correspondingly, lots of research in psychology and other fields has investigated relationship breakups, looking at dimensions such as the impact of breakups on mental health \cite{rhoades2011breaking}, post-breakup personal growth \cite{PERE039}, or the increased use of technology for the actual act of breaking up \cite{weisskirch2012its}.

Through the advent of social media, it is possible to publicly declare one's relationship either using a dedicated functionality as provided by Facebook's ``relationship status'' or, as in the case of Twitter, stating a relation in one's public profile. For example, @user1 on Twitter might write ``@user2 is the best boyfriend ever!!''. In fact, updating one's social network information to mention a new partner has become almost synonymous with the beginning of a committed relationship, leading to the expression ``Facebook official'' \cite{papp12}.


Given the scale and richness of data available on these social networks, they have proven a treasure trove for studying relationships and relationship breakups. Most of the existing work here that has not relied on small-scale survey data has used proprietary data such as coming from Facebook \cite{backstrom2014romantic} or online dating sites \cite{FioreD05,XiaJWCL14}. 
In this work, we show that it is possible to study relationship breakups using public data from Twitter. Concretely, we analyze data for couples where at least one partner in their profile mentions the other one at the beginning of our study period (Nov. 4, 2013). We then periodically look for removals of this profile mention before Apr. 27, 2014 and take this as indication of a breakup, which we validate using CrowdFlower.

We use this data to address research questions related to 
(i) finding indicators of an imminent breakup in the form of changes in communication patterns, 
(ii) the connection between pre- and post-breakup closeness, 
(iii) evidence for post-breakup depression and its dependence on being either the rejector/rejectee,
and (iv) the connection between ``stonewalling'' and relationship breakups.

Using Twitter data or other public social network data to address such questions comes with a number of advantages, including (i) ease of data collection, (ii) size of data, (iii) less self-reporting bias, (iv) timely collection around the moment of break-up, and (v) having social context in the form of network information. However, using this type of data also comes with a number of drawbacks including (i) noise of data, (ii) lack of well-defined variables, (iii) difficulties in observing psychological variables, (iv) limited power to determine \emph{causal} links, and (v) privacy concerns. We discuss more of the limitations and challenges of our study in Section~\ref{sec:discussion}.

Our findings include:
\begin{itemize}
\item Using crowdsourcing we validate that it is possible to identify a large set of relationship breakups on Twitter.
%
%
\item We observe changes in communication patterns as the breakup approaches, such as a decrease in the fraction of messages to the partner, and an increase in the fraction of messages to other users.
\item We observe batch un-friending and being un-friended as indicated by the sudden loss of both 15-20 Twitter friends\footnote{We use the term ``friend'' as Twitter terminology referring to another Twitter user that a user follows.} and followers.
\item We confirm that couples who breakup tend to be ``fresher'' when compared to couples that do not breakup.
\item We observe an increased usage of ``depressed'' terms after the breakup compared to couples that do not breakup.
\item We find a higher level of depressed term usage for likely ``rejectees'' compared to ``rejectors'', both before and after the breakup.
\item Communication asymmetries, related to one-sided ``stonewalling'', are more likely for couples who will breakup.
\item There are higher levels of post-breakup communication for couples who had higher pre-breakup levels of interaction. 
%
%
\end{itemize}

\section{Data Collection}

Twitter is an online social networking and micro blogging platform. It is one of the biggest social networks with around 270 million active users.\footnote{\url{https://about.twitter.com/company}}.
 Each Twitter user has a profile, also called bio, where they can describe themself. The content of this free text field is referred to as \emph{profile description} in this study.

Terminology -- Messages vs.\ Mentions: We define that a Twitter user @user1 has sent a \emph{message} to @user2 if a tweet by @user1 starts with `@user2'. An example message from @user1 to @user2 could be: ``@user2 can I come over to your place?". These \emph{public} messages are not to be confused with direct messages, which can only be sent to followers, are always private and can not be accessed via the Twitter API. A user @user1 is said to \emph{mention} @user2 in a tweet if @user2 occurs anywhere in the tweet. An example could be ``I love @user2 soo much!".\footnote{\url{https://support.twitter.com/articles/14023-what-are-replies-and-mentions}} Note that all messages as well as all retweets are special kinds of mentions.

Our data collection starts with a 28 hour snapshot of Twitter containing about 80\% of all public tweets in late July 2013 (provided by GNIP\footnote{\url{http://gnip.com/}} as part of a free trial). Each tweet in this data set comes with meta data that includes the user's profile description at the time the tweet was created. We searched this meta data for profiles of users containing mentions of other users and along with terms such as ``boyfriend'', ``girlfriend'', ``love'', ``bf'', ``gf'', or ``taken''. For example, the user profile of @user1 containing ``I am taken by @user2, the love of my life'' would be considered because it mentions another user ``@user2'' and contains the word ``love'' (as well as ``taken''). 
We removed profiles mentioning other accounts of the same person such as on Instagram, Facebook, Vine, etc. by looking at simple word matches like `ig', `instagram', `vine', `fb', etc, or if the user being mentioned is the same as the actual user. e.g. Profiles like ``I love football. Follow me on instagram: @user2'' would be removed.
We also had a few thousands of profiles mentioning popular celebrities, especially @justinbieber and @katyperry. Many of these seemed to indicate one-sided, para-social relationships where people claimed @justinbieber as their ``boyfriend'' or their ``love''.

In the end, we had 78,846 users (39,423 pairs) with at least one of the users in the pair mentioning the other in their profile, tentatively indicating a romantic relationship. We tracked these $\sim$80k users starting from Nov. 4, 2013, till the end of Apr. 2014 (24 weeks). We obtained weekly snapshots of the tweets, user profiles (containing the profile description, their self-declared location, time zone, name, the number of followers/friends/tweets, etc.) and their mutual friendship relations (Does @user1 follow @user2? Does @user2 follow @user1?).
Note that even though we started with a set of $\sim$80k users, some of them deleted their accounts over the course of the study, some of them are private or made them private during the $\sim$6 months of data collection. So by the end of the data collection, we were left with 73,868 users.

For the current study, we limited ourselves to English-speaking countries to avoid cultural differences and difficulties in analyzing different languages, e.g., with respect to sentiment. Hence, we only kept users who had their profile location set to US, Canada or UK, identified using the profile timezone and their profile language set to English. This left us with 6,737 couples (13,474 users).

As our simplistic approach of identifying tentative romantic relationships gave some false positives, such as ``Host for @SacrificeSLife All things video games. I love comic books...'', we used Crowdflower\footnote{\url{http://www.crowdflower.com}} (an online crowd sourcing platform) to clean our data.
Concretely, we asked three human judges to manually label if two users were involved in a romantic relationship in Nov. 2013, and again in Apr. 2014 by looking at the pair of Twitter profile descriptions at the relevant time. So each user couple was labeled for two snapshots in time.

The human judges had to decide on a simple ``Yes/No'' answer, indicating a romantic relationship or not for that snapshot. Since this is a potentially subjective task, the judges were asked to answer ``No'' unless it is very clear that the pair are in a romantic relationship. To ensure additional quality, we only used results where all the three human judges agreed on a label. All three judges agreed on the same label in 66\% of the cases.

From this labeling, we can infer if a couple who were in a romantic relationship at the start of the study (Nov. 04, 2013), were still in a relationship at the end of the study (27 Apr. 2014). If they were in a relationship in Nov. and not in Apr., we assume that the couple broke up sometime during this period.

We also used Crowdflower to manually label the gender of the users given the name, profile description and profile picture, again using three human judges for each task. The judges had to pick one of ``Male/Female/Cannot say'' about the gender of the Twitter profile. Similar to the above task, we made sure that the labels were of good quality and picked only those users for whom all the judges agreed on a gender (80\% inter-judge agreement). We also ignored the users who were labeled ``Cannot say''.

In the end, we obtained 1,250 pairs of users highly likely to have undergone a relationship breakup, as well as 2,301 pairs of users who were in a genuine romantic relationship but did not breakup. Para-social relationships with celebrities, mentioned above, were filtered out during this step as the \emph{pair} was not labeled to be in a romantic relationship to begin with.

We decided to remove couples likely to be married, using a simple keyword search for ``married'', ``wife'', ``husband'', etc., as these groups have been observed to follow different relationship dynamics compared to casual dating relationships \cite{cupach1986accounts}. There were also a small fraction of same-sex couples which were removed as, again, they are likely to follow other dynamics \cite{gottman2003correlates}.
This left us with a set of 661 pairs (1,322 users) which we refer to as BR. As a reference set, we also randomly sampled a set of 661 pairs of users who we knew were in a romantic relationship, but did not breakup over the course of our study. We refer to this set as NBR.

\begin{figure}
  \begin{subfigure}[b]{0.5\textwidth}
	\includegraphics[width=\textwidth, clip=true, trim=5 0 5 15]{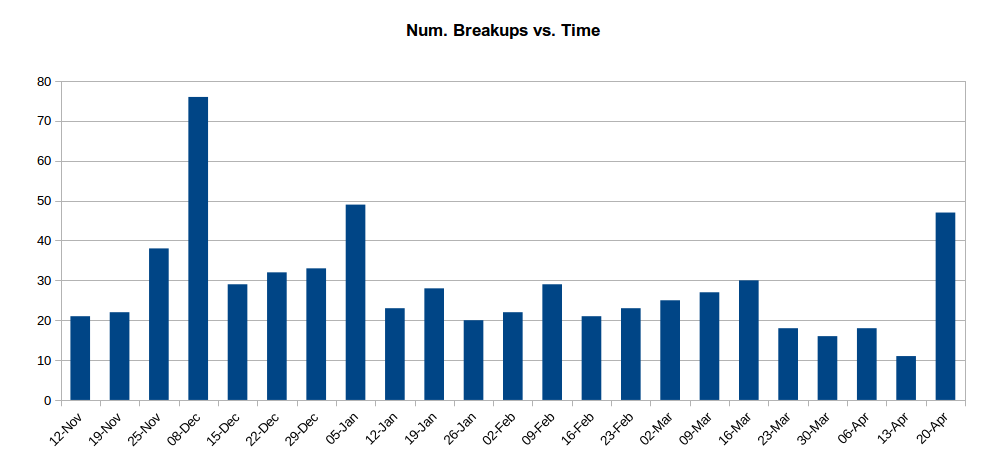}
	\caption{Number of breakups in our data set over time.}
	\label{fig:num_breakups_time}
  \end{subfigure}
  \begin{subfigure}[b]{0.5\textwidth}
	\includegraphics[width=\textwidth, clip=true, trim=5 0 5 15]{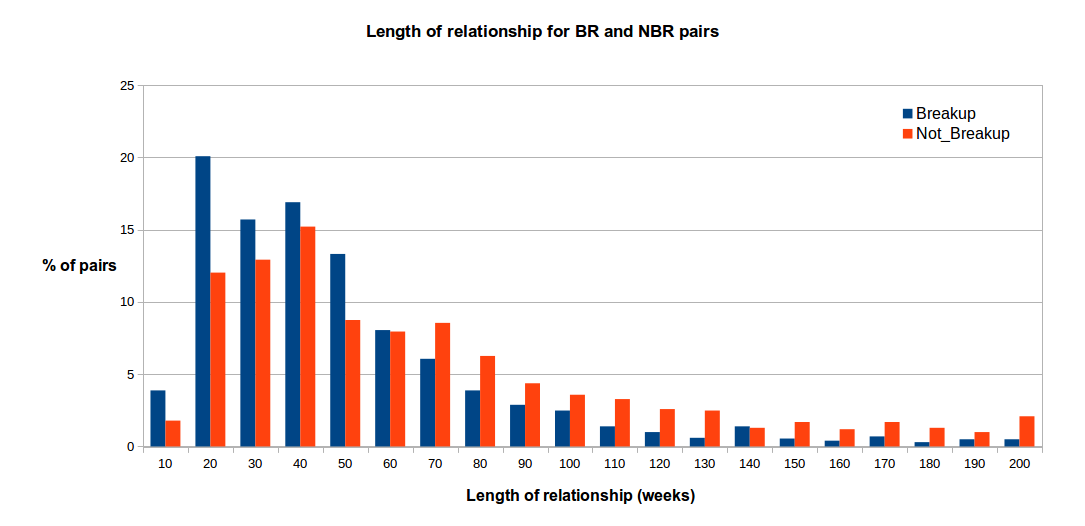}
	\caption{Length of relationships (in weeks) for BR and NBR pairs.}
	\label{fig:relationships_length}
  \end{subfigure}
\end{figure}

For the BR user pairs, we looked at their weekly profile description snapshots and identified the week when at least one user removed the mention of another user in their profile. We define this to be the week the two users broke up. Fig.~\ref{fig:num_breakups_time} shows the distribution of breakups in our data over time. Though there is some temporal variation we did not break down the data further into, say, pre- and post-Christmas breakups.
Still, to avoid temporal-specific peculiarities we also paired the 661 BR pairs with the 661 NBR pairs concerning the week of the breakup. This way we when we refer to ``one week before the breakup'' for a particular couple in our analysis, we use the very \emph{same} week for the randomly paired NBR pair.


\vspace{-1cm}
\section{Results}

\subsection{Length of Relationship}
\label{sec:relationship_length}
It is known from literature that with an increase in relationship duration the breakup probability decreases \cite{le2010predicting}. This is consistent with observations from our data. Concretely, to estimate the length of the relationship between @user1 and @user2, we looked at the oldest tweet in our data set where one mentioned the other. We could do this as, even though we only started our study in Nov. 2013, we collected (up to) 3,200 historic tweets for each user in our study at that point. For the vast majority (80\%) this covered all their tweets.

Using this occurrence of the first mention in a tweet as proxy for relationship duration, we find that the average relationship length for BR pairs in Nov. 2013 was 35 weeks whereas it was 60 weeks for NBR pairs. Figure~\ref{fig:relationships_length} shows a histogram of the estimated relationship duration at the beginning of our study period.

\subsection{Post-Breakup Changes in Profile Description}
\label{sec:word_clouds}

\begin{figure}
\centering
  \begin{subfigure}[b]{0.48\textwidth}
    \includegraphics[width=\textwidth]{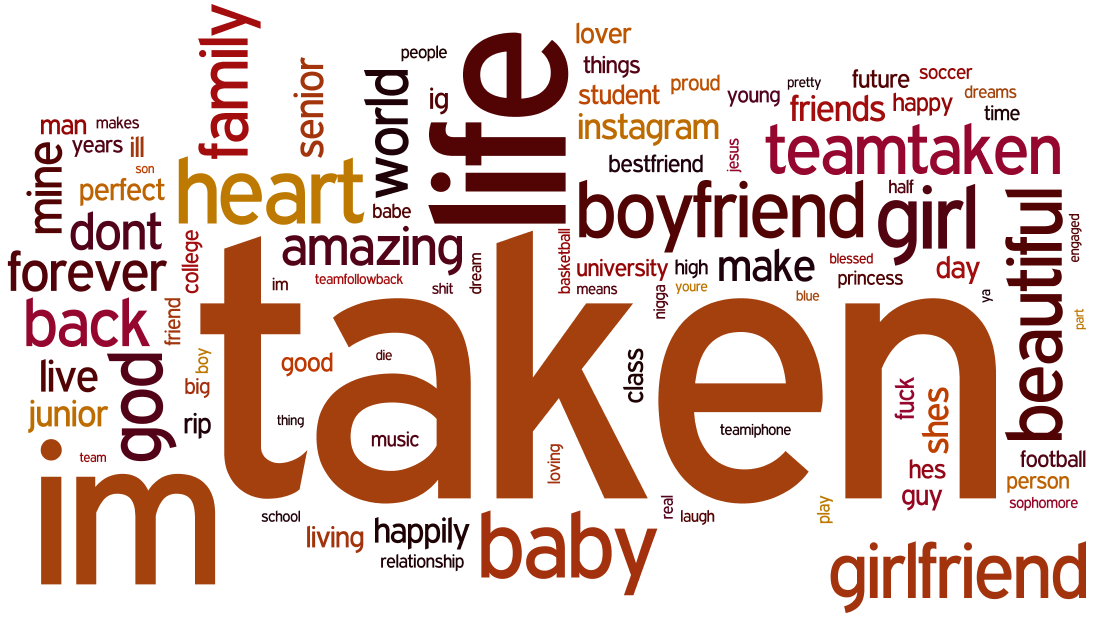}
	\caption{}
    \label{fig:before}
  \end{subfigure}
  \begin{subfigure}[b]{0.48\textwidth}
    \includegraphics[width=\textwidth]{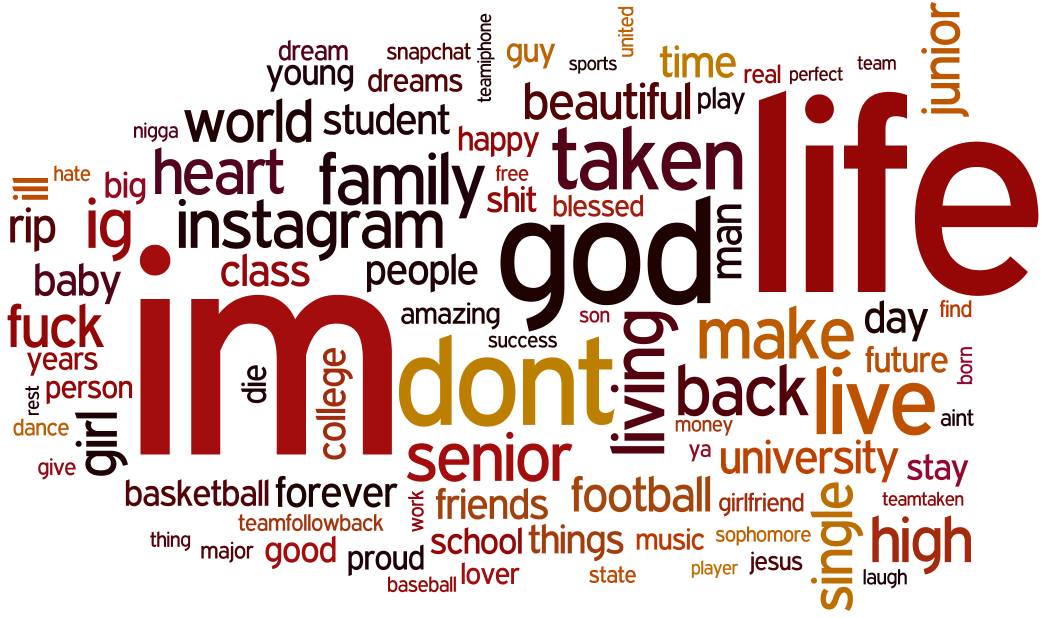}
    \caption{}
    \label{fig:after}
  \end{subfigure}
  \caption{(a) Word cloud of the profile descriptions before breakup, at the beginning of our study. (b) Word cloud of the profile descriptions one week after the breakup.}
\end{figure}

The removal by one user of the mention in their profile description of the other user is, as described before, our definition of a breakup. However, we were interested in which \emph{other} changes of the profile description would coincide with a breakup. To study this, we looked at the profile descriptions of BR users (a) at the start of the data collection - 04 Nov 2013, and (b) the week after their breakup. 
We then generated word clouds for these two sets of profiles. Figures~\ref{fig:before},~\ref{fig:after} show the profiles before/after. (We removed the very frequent words ``love'' and ``follow'' from both before and after as they were at least 75\% more frequent than the next most frequent word before breakup and hence distorting the distribution.)

There are several clearly visible differences and these reconfirm that our data set really does contain actual relationship breakups. For example, the terms ``taken'' or ``baby'' both lose in relative importance compared to the other terms. Note, however, that terms such as ``taken'' do not disappear completely which relates to a temporal difference in when the two partners update their profiles. See Section~\ref{sec:rejector_rejectee} for details.

To rule out the influence of background temporal changes due to, say, Christmas or Valentine's Day we also generated similar word clouds using users from NBR. For this set, we could not observe any differences over time and the figures, very similar to the ``before'' cloud, are omitted here.

Though the relative loss of ``taken'' is expected, we were also interested in which terms would \emph{gain} in relative importance and, in a sense, come to replace the former reference to the partner. To quantify the change in relative importance, we ranked the words before/after by frequency and looked at those words which increased in terms of rank the most. Concretely, we weighted terms by the formula $(before\_rank - after\_rank)/after\_rank$, which gives more weight to terms moving to the top, rather than moving up from, say, rank 100 to rank 80. The top gainers are, in descending order, ``im'', ``god'', ``dont'', ``live'', ``single'', ``dreams'', ``blessed'', ``fuck''.
One story that potentially emerges from this is that people (i) become more self-centered, (ii) find stability in religion and spirituality, but also (iii) curse life for what has happened. A positive impact of spirituality on post-breakup coping has also been observed before \cite{hawley2012roles}.

\vspace{-0.5cm}
\subsection{Changes in Communication Styles with (Ex-)Partner}
\label{sec:4grams}

\begin{figure}
\centering
  \begin{subfigure}[b]{0.48\textwidth}
    \includegraphics[width=\textwidth]{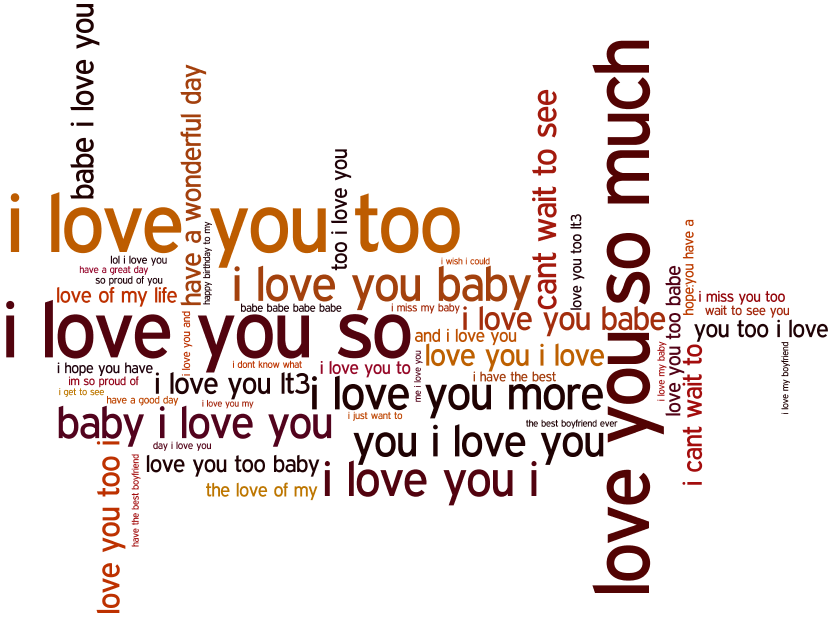}
    \caption{}
    \label{fig:before_messages_4_grams}
  \end{subfigure}
  \begin{subfigure}[b]{0.48\textwidth}
    \includegraphics[width=\textwidth]{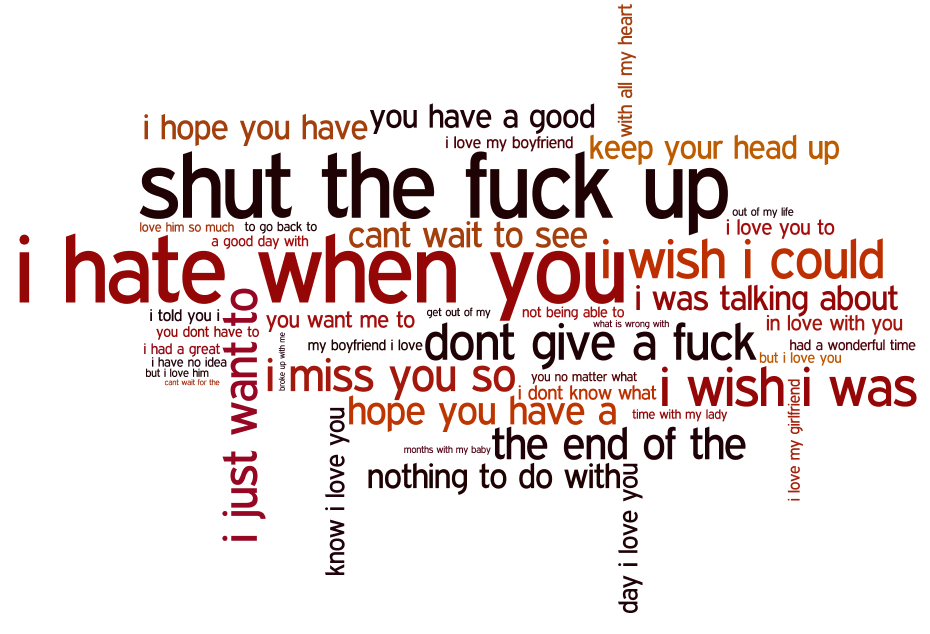}
    \caption{}
    \label{fig:after_messages_4_grams}
  \end{subfigure}
  \caption{(a) Word cloud of the 4-grams from messages exchanged between BR users before breakup. (b) Word cloud of the 4-grams from messages exchanged between BR users after breakup.}
\end{figure}

As Twitter is used for many purposes, including sharing factual information, we were interested to see if there would be a noticeable change in tone when one partner would message the other, either before or after the breakup. As simple analysis tools, we generated word clouds of 4-grams of words from conversations (messages) between pairs of users breaking up. Figures~\ref{fig:before_messages_4_grams} and \ref{fig:after_messages_4_grams} show the shifts in personal communication patterns. 

The change is roughly from ``I love you so ...'' to ``I hate when you ...'', indicating a (to us) surprising amount of \emph{public} fighting and insulting happening after the breakup. For future analysis it might be interesting to quantify which relationships ``turn sour'', e.g., as a function of pre-breakup closeness.

\vspace{-1.2cm}
\subsection{Changes in Communication Patterns Around Breakups}
\label{sec:communication_changes}

Apart from looking for expected before/after changes, we were interested to see if there were any gradual changes in communication patterns as people gradually edged towards a breakup.
For this, we considered only those users who had at least four weeks of data before and two weeks after the breakup (1,070 users). We then looked at changes in (i) the fraction of tweets that contain a message to the partner, (ii) the fraction of tweets that are messages to non-partner users, and (iii) the fraction of tweets that are ``original'', i.e., non-retweet tweets.
In all cases, these were then macro-averaged such that each couple, independent of their number of tweets, contributed equally to the average.

\begin{figure}
\centering
  \begin{subfigure}[b]{0.33\textwidth}
    \includegraphics[width=\textwidth]{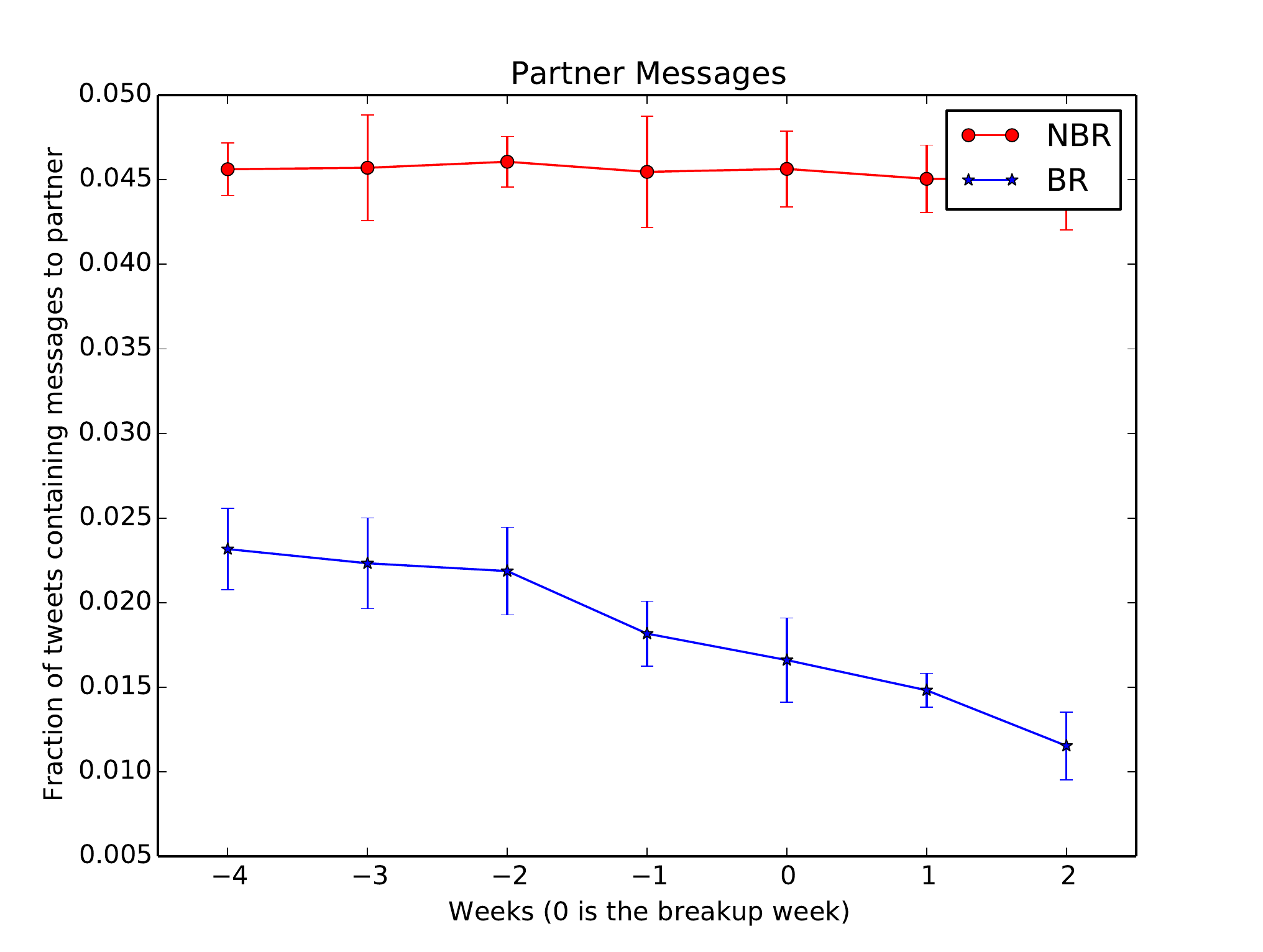}
    \caption{}
    \label{fig:before_after_frac_num_mentions_partner}
  \end{subfigure}
    \begin{subfigure}[b]{0.32\textwidth}
    \includegraphics[width=\textwidth]{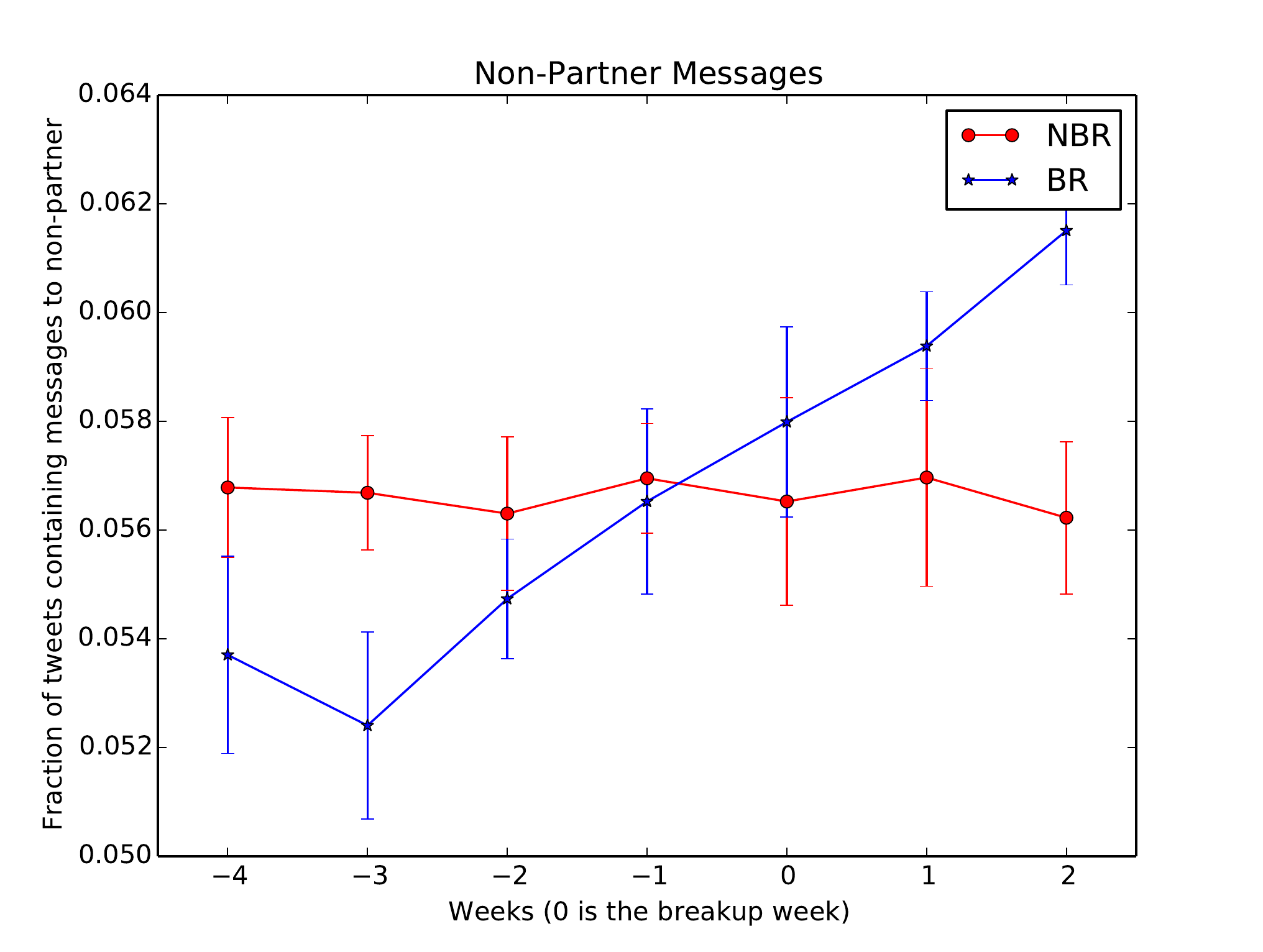}
    \caption{}
    \label{fig:before_after_frac_num_messages_non_partner}
  \end{subfigure}
    \begin{subfigure}[b]{0.33\textwidth}
    \includegraphics[width=\textwidth]{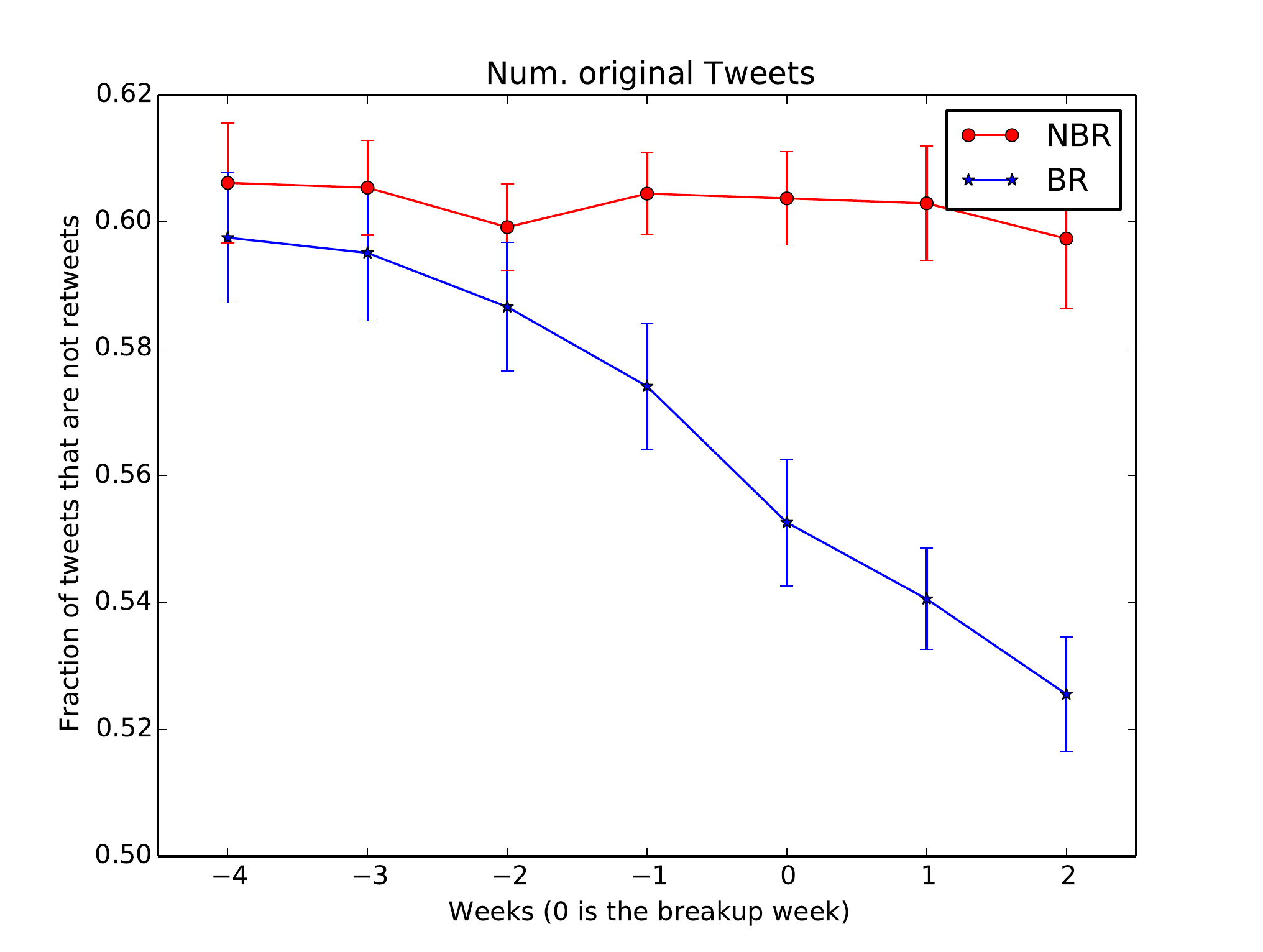}
    \caption{}
    \label{fig:before_after_frac_num_original_tweets}
  \end{subfigure}
  \caption{Comparison of various features using data from four weeks before, during and two weeks after the breakup. (a) Fraction of the total tweets containing mentions of the partner. (b) Fraction of the total tweets containing direct messages to someone other than the partner. (c) Fraction of the total tweets that are not retweets. Error bars indicate standard errors.}
\end{figure}

The trends are noticeable and consistent: as the breakup approaches -- and beyond -- (i) the number of messages to the partner decreases, (ii) the number of messages to other users increases, and (iii) the fraction of original tweets goes down. Though not the goal of this study, these observations could potentially lead to ``early breakup warning'' systems.

\vspace{-0.3cm}
\subsection{Breakup-induced Batch Un-Friending and Being Un-Friended}
\label{sec:unfriending}
After the breakup, we were expecting partners to potentially unfollow each other but, apart from that, we were expecting ``business as usual'' as far as the social network was concerned. However, when we tried to quantify our hypothesis that there should not be ripple effects affecting other connections, we found evidence for the opposite.

Concretely, we monitored the number of friends and followers of each of our users over time. To be able to quantify the temporal changes, we only considered users who had at least two weeks of data before and two weeks after the breakup (1,156 users). As can be seen in Table~\ref{tab:num_friends}, for the BR pairs there is a sudden drop of about 20 followers/friends on average and 16 in the median. 
Unfortunately, we do not have data on \emph{who} is being unfollowed or stops following and we can only speculate that these are former mutual friends (We had to remove an outlier user @tatteddarkskin (249k followers, 270k friends), who changed this Twitter id to @iammald, the week he broke up).

\begin{table}
\centering
\begin{tabular}{c|c|c|c|c|c}
& & & friends & & \\
\hline
& T-2 & T-1 & T0 & T+1 & T+2 \\
\hline
BR & 579 (294) & 582 (295) & 562 (280) & 564 (281) & 577 (285) \\
NBR & 588 (273) & 591 (273) & 596 (275) & 598 (275) & 601 (276) \\
\hline
& & & followers & & \\
\hline
BR & 683 (328) & 689 (329) & 669 (313) & 672 (314) & 675 (316) \\
NBR & 778 (285) & 780 (288) & 785 (290) & 787 (291) & 788 (292) \\
\end{tabular}
\caption{Average number of friends/followers for BR and NBR for two weeks before and two weeks after the breakup. The numbers in parentheses are for the median.}
\label{tab:num_friends}
\end{table}

%
%
%

\vspace{-1.3cm}
\subsection{Making Profiles Private}
\label{sec:private}
Given that relationship breakups can be traumatic experiences and that going through this in public can potentially be perceived as embarrassing, we wanted to see if breakups have effects on users' privacy settings.

On Twitter, by default all information is public and anyone can read your tweets and see your network information. However, Twitter users have the option to make their account private, which restricts access to their tweets to their followers where each follower now requires approval by the user. However, even for private profiles, the profile description and meta information such as the number of tweets, friends or followers remains accessible through the Twitter API. But the tweets' content or the identities of users in the user's social network are then hidden.

Out of 1,250 users that ever broke up, (excluding users who broke up in the very first and last week), 98 users eventually made their account private. Of these, 74 users made their account private within +/- one week of the breakup with 22 users already making this change \emph{before} the week of the breakup.
On the other hand, only 23 of the NBR users made their account private.

Put differently, BR users had a 7\% probability of making their profiles private whereas this was 2\% for NBR users (excluding first and last week).
Interestingly, though there is work on privacy issues on Twitter \cite{Jin13,MaoSK11}, we are not aware of any study that looks at when and why users change their Twitter privacy settings. For Facebook on the other hand, a connection between relationship breakups and changes in privacy settings has been observed \cite{lefebvre2014navigating}.

\vspace{-0.4cm}
\subsection{Evidence for Post-Breakup Depression}
\label{sec:depression}
Relationship breakups are known to be linked to depression \cite{sbarra2005emotional}. As far as Twitter is concerned it has also been shown that tweets can give indications of depression \cite{de2013predicting,TsugawaMKKFIO13}.
Following features used as part of depression classifiers, we decided to use certain categories of the Linguistic Inquire and Word Count (LIWC) dictionary\footnote{\url{http://www.liwc.net/}}~\cite{pennebaker2007development}.
Concretely, we combined terms from the ``sad'', ``negemo'' (= negative emotions), ``anger'' and ``anxious'' categories into a single category we call ``depressed''. The choice of these categories to be merged is inspired by \cite{tong2013facebook} who find that 
 ``much of the research on uncertainty reduction theory (URT) has documented that high levels of uncertainty between romantic partners are correlated with greater feelings of anger, sadness, and fear, and that reduced uncertainty is accompanied by a decrease in the experience of negative emotion.''

For the ``depressed'' category we then looked at the fraction of tweets during a particular week that contained at least one term from this category. These fractions were then first averaged for each partner of a couple and then averaged across couples. The resulting fractions over time are shown in Table~\ref{tab:depression}.

\begin{table}
\centering
\begin{tabular}{c|c|c|c|c|c}
\hline
& T-2 & T-1 & T0 & T+1 & T+2 \\
\hline
BR & 0.124 & 0.129 & 0.132 & 0.143 & 0.149 \\
NBR & 0.105 & 0.106 & 0.107 & 0.104 & 0.107 \\
\hline
\end{tabular}
\caption{Fraction of tweets containing depressing words, 2 weeks before and after the breakup.}
\label{tab:depression}
\end{table}

For each week we find a statistically significant difference between BR and NBR couples ($p<.01$ using non-parametric bootstrap resampling) where BR pairs consistently have higher levels of these words. Moreover, their usage of these terms increases over time and the difference between T-2 and T+2 are significant ($p<.01$ using non-parametric bootstrap resampling).

\vspace{-0.3cm}
\subsection{Being Dumped Hurts More Than Dumping}
\label{sec:rejector_rejectee}
When it comes to coping with relationship breakups previous work has found differences depending on whether a person is the ``rejector'', i.e., the initiator of the break-up or the ``rejectee'' \cite{perilloux2008breaking,sprecher1998factors}.
To identify potential breakup initiators, we looked at BR pairs that initially had a mutual profile reference, but where one removed the mention of the other earlier, i.e., not in the same week. We hypothesize that the initiators are first to remove the reference of the former partner and label them ``Rejectors'' and the others were ``Rejectees''. There were 164 pairs where we observed such a behavior. Out of these, in 67\% of the cases, women were the rejectors.


As far as word usage of the ``depressed'' terms is concerned, we found that rejectors feel less depressed compared to the rejectees (observed previously in \cite{perilloux2008breaking} and \cite{sprecher1998factors}) as shown in Table~\ref{tab:initiators_depression}. Again, the differences between pairs in the same week and the weeks T-2 and T+2 were tested for significance using bootstrap sampling and found to be significant at $p<.01$.

\begin{table}
\centering
\begin{tabular}{c|c|c|c|c|c}
\hline
& T-2 & T-1 & T0 & T+1 & T+2 \\
\hline

Rejector & 0.116 & 0.124 & 0.125 & 0.129 & 0.131 \\
Rejectee & 0.138 & 0.128 & 0.145 & 0.154 & 0.163 \\
\hline
\end{tabular}
\caption{Rejector's vs.\ rejectee's depression levels before and after breakup.}
\label{tab:initiators_depression}
\end{table}

%

\vspace{-1.5cm}
\subsection{Pre-Breakup Communication Asymmetries}
\label{sec:stonewalling}

Stonewalling is one of the ``four horsemen of the apocalypse'' defined by Gottman~\cite{gottman1995marriages}. Stonewalling refers to ignoring the other partner and we quantify it by looking for communication asymmetries, where if only one side is ``doing all the talking'' there is evidence of stonewalling. 
Concretely, in each of the four weeks before the breakup, we looked at the number of messages exchanged between users. Here we looked if there were at least five times as many messages in one direction as the other direction and an absolute difference of at least five messages (to avoid cases where the difference was a mere one or two messages vs.\ zero messages). 
For BR couples, we found this kind of stonewalling in 224 out of the 585 couples (38\%). For NBR, we only found it in 59 out of the 585 couples (10\%).

\vspace{-0.4cm}
\subsection{Post-Breakup Closeness}
\label{sec:post_breakup_closeness}

Existing work has looked at predicting post-breakup closeness using pre-breakup closeness \cite{tan2014committed} and found a positive connection between the two. We show that our data set can also be used to study this phenomenon by operationalizing these concepts as follows. We mark a pair as being ``close'' after a breakup if they both mention each other at least in two distinct weeks after they breakup (requiring a total of at least four tweets).
97 BR couples (16\% of all BR couples with at least two post-breakup weeks) satisfy this condition for maintained, bi-directional communication and we call them PBC (for post-breakup closeness).

To quantify pre-breakup closeness, we looked at a user's fraction of all pre-breakup tweets that were messages to the partner. We did the same for their partner and then averaged this value for this couple, and then across all couples. This we did separately for the PBC set and the remaining BR pairs called NPBC. The same procedure of averaging pre-breakup tweets ratios was repeated for (i) the fraction of mentions to the partner and (ii) the fraction of retweets of the partner. We also obtained (up to) 3,200 of a user's favorites at the end of the study period and looked at the fraction of those that were for tweets by the partner. This value was again averaged across partners and then across couples.
The results are presented in Table~\ref{tab:post_breakup_closeness}. For all four measures of ``closeness'' there is a significant difference between the PBC and the NPBC groups with higher levels of pre-breakup communication and interaction for couples who stay in touch after the breakup, confirming results in~\cite{tan2014committed}. We also found the same trend when looking at the pre-study relationship duration (see Section~\ref{sec:relationship_length}) and the average relationship duration was 47 weeks for PBC, but only 34 for NPBC.

\begin{table}
\centering
\begin{tabular}{|c|c|c|c|c|c|}
\hline
& Messages & Mentions & Retweets & Favorites \\
\hline
PBC & 0.0559 & 0.0842 & 0.011 & 0.0897 \\
NPBC & 0.0296 & 0.0551 & 0.006 & 0.0505 \\
\hline
\end{tabular}
\caption{Difference between various interaction related features for PBC and NPBC. All PBC and NPBC values are statistically significantly different ($p<.01$ using bootstrap resampling).}
\label{tab:post_breakup_closeness}
\end{table}


\vspace{-1.5cm}
\section{Related Work}
\vspace{-0.2cm}

The only work we know of on studying romantic relationships on Twitter is Clayton et al.\ \cite{clayton2014third}. Using answers to specific questions (from surveys) from a few hundred users, they look at how Twitter mediates conflict between couples. They find evidence that ``active Twitter use leads to greater amounts of Twitter-related conflict among romantic partners, which in turn leads to infidelity, breakup, and divorce''.

Currently, we are using Twitter merely as a data source to study relationship breakups per se. However, one could also study the more intricate relationship between technology use and personal relationships.
Weisskirch, et al.\ \cite{weisskirch2012its} look at the attachment styles of couples involved in a relationship breakup online. It is the only work that we are aware of that looks at the act of breaking up through technology. Manual inspection of tweets around breakup revealed a few instances of actual breakups through public (!) tweets in our dataset too.

Apart from facilitating breakups, increased importance of technology in romantic relationships \cite{papp12} potentially has other negative impact on romantic relationships such as jealousy, or surveillance \cite{tong2013facebook,ClaytonNS13,DrouinaM14}. On the positive side, researchers have looked at if technologies such as video chat can positively affect long-distance relationships by making it easier to feel connected \cite{Hassenzahl2012,Neustaedter2012}.

Hogerbrugge et al.\ \cite{hogerbrugge2013dissolving} study the importance of social networks in the dissolution of a romantic relationship. They define certain factors such as the overlap of networks of partners or social capital and study how these factors affect breakup. Though we did not collect data for the Twitter social \emph{network}, or its changes over time, it would be possible to validate their findings on Twitter using our approach of identifying breakups.

Backstrom et al.~\cite{backstrom2014romantic} used the network structure of an individual's ego network to identify their romantic partner. Note that a social tie on Facebook is not the same as one on Twitter, mainly because, (i) Twitter network is directed, (ii) the use of Facebook and Twitter may be different. Still, the notion of `dispersion' defined in their paper might be related to the loss of friends/followers in our study (see Section~\ref{sec:unfriending}).
Lefebvre et al.\ \cite{lefebvre2014navigating} study relationship dissolution on Facebook, mainly focusing on the phases and behavior of users who go through breakups on Facebook. There is evidence of limiting profile access in order to manage the breakup which is similar to our findings in Section~\ref{sec:private}.

Researchers conducting retrospective~\cite{frazier1993correlates} and
diary~\cite{sbarra2005emotional} studies of emotional adjustment
following a breakup have found evidence of negative emotional
responses including sadness and anger. In contrast to the current
findings, Sbarra et al. found no difference between rejectors and
rejectees in the extent of negative emotion following a breakup, and
suggested that this might reflect difficulties in accurately
identifying who initiated the breakup. Though imperfect, the current
approach of identifying the first person to remove a profile mention
as the ``initiator'' or ``rejector'' may provide a good proxy for being
the person who is more ready to terminate the relationship or who
feels more control over the breakup; this latter feature of
controllability has been found to predict better adjustment post
breakup \cite{frazier1993correlates,sprecher1998factors}.
It may also be that the larger sample size in the current study
provides more statistical power to detect these effects than has been
available in smaller survey studies.

Researchers studying close relationships have identified a number of
factors that predict longevity and dissolution of non-marital
relationships, including duration of the relationship, commitment,
closeness, conflict, inclusion of other in the self, and the
availability of alternatives~\cite{le2010predicting}. Our analyses were informed
by these extant findings, and we attempted to identify proxies for
several of these important predictors, e,g., that with bi-directional
profile mentions might be a sign of greater commitment than
unidirectional mentions, and our findings tend to support those of the
meta-analysis. However, we note that some relationship features may be
more easily extracted from Twitter data than others. Factors like
duration of the relationship and conflict might emerge clearly in the
Twitter data (e.g., Figures~\ref{fig:relationships_length}, \ref{fig:after_messages_4_grams}). Others, such as commitment or
inclusion of other in the self (IOS), are typically assessed using
multi-item self-report questionnaires, and are not directly observable
in tweets (at least not with any degree of frequency). Therefore,
computational social scientists should pay particular attention to the
need for studies that demonstrate the relations between the patterns
they observe in data from online social networks and validated
measures of relationship factors. For e.g., in a study of romantic
relationships on Facebook~\cite{carpenter2013exploring} , participants' IOS
scores were related to frequency of tagging (in status updates and
in photos) but not to the number of mutual friends or
proportion of shared interests (both of which could be plausible
correlates of IOS).

\vspace{-0.5cm}
\section{Discussion}
\label{sec:discussion}
\vspace{-0.3cm}



Though our point of departure was a privileged data set, derived from a trial period for data access by GNIP, other ways to gather data are possible. For example, one could use services such as Followerwonk\footnote{\url{http://followerwonk.com/bio/}} to obtain a list of Twitter users with ``boyfriend'' in their profile description. For historic studies, one could use from the 1\% ``Spritzer'' sample of public tweets on the Internet Archive\footnote{\url{https://archive.org/details/twitterstream}} to find a sample of such users.

As with most similar, observational studies reasoning about causal links is tricky. For example, the increased usage of depressed terms (see Section~\ref{sec:depression}) after the breakup could be a consequence of the breakup itself, or it could indicate that relationships are more likely to end when someone is about to undergo increased levels of depression. 

Despite being limited when it comes to detecting causal links, observational studies such as ours are useful to validate existing models and theories as well as to provider pointers as to where a more in-depth study could be promising.
For example, the observation that there is a sudden loss both in the number of friends and followers (see Section~\ref{sec:unfriending}) is worth following up on. Were those to-be-removed friends only added due to social pressure in the first place? Or were they actual ``friends'' but maintaining communication with them would have been too emotionally taxing?

Most existing work on post relationship breakup behavior is based on surveys conducted long after the breakup, where people typically recall the experiences they have been through. This method has serious flaws as pointed out in~\cite{sakaluk2013problems}. Fortunately, we can collect data right around the time of the breakup.

Undoubtedly, couples in our data set are \emph{not} representative of all heterosexual dating relationships in the United States, UK and Canada. Manually inspecting the data indicates an over-representation of teenagers.
However, even the set of teenage dating relationships make up a significant part of relationships and are worth studying, especially as they seem to follow established patterns when it comes to the effect of the duration of relationship on the breakup probability (c.f.\ Section~\ref{sec:relationship_length} and \cite{le2010predicting}), the occurrence of post-breakup depression (c.f.\ Section~\ref{sec:depression} and \cite{sbarra2005emotional}), or communication asymmetries in the form of ``stonewalling'' (c.f.\ Section~\ref{sec:stonewalling} and \cite{gottman1995marriages}).

Even though in this study we ignored para-social relationships with teen celebrities like @justinbieber and @katyperry, we could have looked at how these relationships evolve over time~\cite{cohen2004parasocial}. Maybe a ``breakup'' with Justin Bieber exerts just as much emotional stress as a breakup with a real boyfriend.

For the current study, we only looked at one-time relationship dissolutions. We did not try to identify cases where a couple got together again or cases where a partner ``moved on'' and entered a new relationship (even though the latter is easy to identify from our data set). Having a larger and periodically updated set of couples to monitor could allow studying studying such phenomena as well.

Arguably, Twitter could be used more as a type of ``information network'' than a ``social network'' \cite{KwakLPM10}, questioning its use as a data source for interpersonal relations. However, Myers et al.\  found that ``from an individual user's perspective, Twitter starts off more like an information network, but evolves to behave more like a social network''. Also, in our study we only consider people who at least partly use it as a social network for personal relations in the first place.

For this study we built a data set with a ``high precision'' approach, at the potential expense of recall. To be considered a ``couple in a relationship'', each pair of users underwent a sequence of filtering steps, including crowd labeling. The scale of our study could be improved by turning to machine-learned classifiers to detect romantic relationships even when partners are not mentioning each other in their profile descriptions. This is similar to work that looks into ``when a friend in Twitter is a friend in life'' \cite{XieLZLG12} and work that classifies pairs of communicating users on Twitter into friends or foes \cite{LiuW14}.



So far we have only looked at basic measures of communication styles, such as the fraction of tweets mentioning a partner. However, there has been a body of work on inferring \emph{personality traits} from Twitter usage \cite{QuerciaKSC11,HughesRBL12,SumnerBBP12,GolbeckRET11}. This work could potentially be applied to our data set to look more into which types of personalities undergo what types of relationship breakups.

Not focusing on romantic relationships, there is research looking at unfollowing on Twitter \cite{Moon11,Kivran-SwaineGN11,KwakCM11,KwakML12}. Though unfollowing could be seen as a ``mini-breakup'' we observed that, maybe surprisingly, 44\% of couples still follow each other two weeks after the week of the breakup and in another 32\% of cases one of the partners still follows the other one at this point. For comparison, initially 96\% of couples mutually follow each other.




\vspace{-0.3cm}
\section{Conclusions}
\vspace{-0.2cm}

We used public Twitter data  to analyze the dissolution of 661 romantic relationships on Twitter during the period of Nov. 2013 to Apr. 2014. We compared the behavior of the users involved with those of 661 couples to those that did not breakup during the same period.
Our analysis confirmed a number of existing hypotheses such as:
(i) the breakup probability decreases with length of the relationship, (ii) post-breakup usage of ``depressed'' terms increases, (iii) rejectees have higher levels of usage of depressed terms compared to rejectors, (iv) communication asymmetries and one-sided stonewalling is indicative of breakups, and (v) higher levels of post-breakup closeness for couples who also have a higher pre-breakup closeness.

We also found evidence of the, to our knowledge, undocumented phenomenon of ``batch un-friending and being un-friended'' at the end of a relationship. Concretely, we observed sudden drops of size 15-20 for both the number of friends and followers a user has around the time of the breakup.

Though our data set is undoubtedly not representative of all relationship breakups we believe our study still shows the huge potential that public social media offers with respect to studying sociological and psychological processes in a scalable and non-obtrusive manner.

%
%

\bibliographystyle{splncs03}
\bibliography{twitter_breakup}

\end{document}